\documentclass[12pt]{article}
\usepackage{graphicx}
\usepackage{amsmath}
\newcommand{\rcorr}{\hbox{\kern-1.2em$\longrightarrow$}}
\newcommand{\lrcorr}{\hbox{\kern-1.2em$\longleftrightarrow$}}

\newcommand{\nRightarrow}{\Rightarrow\kern-1.2em\hbox{/}\kern.8em} %
\newcommand{\BB}{\hbox{I\kern-.2em\hbox{B}}} %BB
\newcommand{\DD}{\hbox{I\kern-.2em\hbox{D}}} %DD
\newcommand{\FF}{\hbox{I\kern-.2em\hbox{F}}} %FF
\newcommand{\NN}{\hbox{I\kern-.2em\hbox{N}}}  %Naturali
\newcommand{\ZZ}{{{\rm Z}\kern-.28em{\rm Z}}} %Interi
\newcommand{\RR}{\mathop{{\rm I}\kern-.2em{\rm R}}\nolimits} %Reali
\newcommand{\QQ}{\hbox{l\kern-.36em\hbox{Q}}}  %Razionali
\newcommand{\CC}{\hbox{I\kern-.58em\hbox{C}}}

\begin{document}
\title{``Evaluations'' of Observables Versus Measurements in Quantum Theory}
\author{Giuseppe Nistic\`o and Angela Sestito\footnote{Part of Angela Sestito's work is
supported  by the European Commission, European Social Fund and  by
the Calabria Region, Regional Operative Program (ROP) Calabria ESF
2007/2013 - IV Axis Human Capital - Operative Objective M2- Action
D.5 }\\{\small Dipartimento
di Matematica, Universit\`a della Calabria, Italy}\\
{\small and}\\
{\small
INFN -- gruppo collegato di Cosenza, Italy}\\{\small email: gnistico@unical.it}}
\maketitle
\abstract{In Quantum Physics there are circumstances where the direct measurement of
particular observables encounters difficulties; in some of these
cases, however, its value can be {\sl evaluated}, i.e. it can be
inferred by measuring {\sl another} observable characterized by
perfect correlation with the observable of interest. Though an {\sl
evaluation} is often interpreted as a {\sl measurement} of the
evaluated observable, we prove that the two concepts cannot be
identified in Quantum Physics, because the identification yields
contradictions. Then, we establish
the conceptual status of evaluations in Quantum Theory
and the role can be ascribed to them.
}
%%%%%%%%%%%%%%%%%%%%%%%%%%%%%%%%%%%%%%%%%%%%%%%%%%%%%%%%%%%%%%%%%%%%%%%%%%%%%%%%%%%%%
\section{Introduction}
In Quantum Physics there are circumstances where some difficulties
encountered in measuring observables are outflanked by exploiting
the correlations existing among observables.
\par
As an example, we can consider a typical Stern\&Gerlach experiment for
a spin-1/2 particle, where the gradient of the magnetic field is oriented along $z$.
The $z$ component $S_z$ of the spin is an observable which pertains to an {\sl internal}
degree of freedom of the particle, so that it is difficult to concretely design a direct
measurement of $S_z$. However, the observable
$T_{up}$ which localizes the particle in the upper exit of the magnet is perfectly
correlated, according to the laws of Quantum Mechanics,
with the values of $S_z$; hence,
the value $+1/2$ of $S_z$ is inferred for the out-coming particles
localized in the upper exit by a measurement of $T_{up}$.
\par
Such a method is used also to outflank the obstacles raised by the fundamental principles of
Quantum Mechanics.
For instance, in a typical double-slit experiment the observable $W$ that indicates
the slit the particle passes through is represented
by a localization operator $\hat W$ which {\it does not commute} with the operator $\hat Q_F$ representing
the impact position on the final screen,
because $W$ and $Q_F$ are positions at different times.
Therefore, the measurement of $W$ is {\sl in principle}
forbidden for a particle whose final position $Q_F$ is measured.
However, under suitable conditions \cite{1}, an observable $T_W$ exists
such that $[\hat T_W,\hat Q_F]={\bf 0}$, and whose outcomes are
perfectly correlated with the outcomes of $W$ in every
simultaneous measurement of $T_W$ and $W$; so, by measuring
$T_W$ and $Q_F$ together,
which slit information is inferred
from the outcome of $T_W$, {\it via} the perfect correlation
between $T_W$ and $W$ without measuring $W$, while $Q_F$ is directly measured.
\par
{\it A priori},
{\sl to assign an observable $E$ the value obtained
as actual outcome of a correlated observable $T$} should be distinguished from a {\sl direct
measurement of $E$}; so, we call it {\it evaluation} of $E$ by $T$.
The different statuses of {\it measurement} and {\it evaluation} in Quantum Theory
are established in section 2.1 and 2.2.
\par
In this work we address the problem
of establishing to what extent evaluations can be interpreted as real measurements.
In fact, in the
experiment of Stern\&Gerlach, for instance, the localization of the particle by $T_{up}$ is interpreted as a
valid measurement of the spin $S_z$.
But, from a theoretical point of view the problem exists.
Indeed in section 2.3 we show that to identify the evaluation
of an observable with its measurement {\sl leads to contradictions} in Quantum Mechanics.
\par
Now, since evaluations are diffusely practiced in Quantum Physics, the
task of establishing how they are related to measurements
cannot be overlooked in Quantum Theory.
\par
Then, in section 3 we shows that evaluations behave as {\sl perfect
simulations} of the measurement of the evaluated observable; more precisely, we show that
the physical consequences of the occurrence of every measurement's outcome of an observable
are physically indistinguishable from the
consequences of the occurrence of the same outcome for its evaluation.
\par
However, in section 4 we point out that the interpretation of evaluations as simulations does not
apply if the evaluation of an observable $E$ by $T$ is performed together with the measurement of
another observable $F$ which does not commute with the evaluated
observable $E$. %; an example is just our two slit experiment, where {\sl
%which slit} observable is evaluated by $T_S$ simultaneously to the
%measurement of $Q_F$. A measurement of the evaluated observable could
%not even happen, and then there would be nothing to simulate.
In this more general case we prove that
a unique joint probability $p_\rho(E\& F)$ exists which rules over a value assignment
for $E$ consistent with the simultaneous occurrence of outcomes of actually performed
measurements of $F$; furthermore, we prove that just to assign $E$ the values evaluated by $T$
realizes such a unique probability. As a consequence, though
$E$ does not commute with $F$, the
evaluated observable $E$ can be assigned the value evaluated by $T$
without violating logical consistency and such an assignment
consistently extends the actual measurements' outcomes.
\par
The impossibility of identifying evaluations with
measurement proved in section 2 is then explained on the basis of our results.
%%%%%%%%%%%%%%%%%%%%%%%%%%%%%%%%%%%%%%%%%%%%%%%%%%%%%%%%%%%%%%%%%%%%%%%%%%%%%%%%%%%%%%%%%%%%%%%%%
\section{``Evaluations'' in Quantum Theory}
Here, in section 2.2, we formally establish the concept of {\sl evaluation} within Quantum Theory.
To do this, in section 2.1 we have to
make explicit some implications of the standard interpretation of Quantum Theory.
Section 2.3 shows that evaluations cannot be identified with measurements,
because the identifications provokes contradictions in Quantum Mechanics.
\subsection{Basic Formalism}
Let $\mathcal H$ be the Hilbert space of the quantum theory of
the investigated physical system. Given any observable $A$, let  $\hat A$ be the corresponding self-adjoint operator,
the expected
value of $A$ is $Tr(\rho\hat A)$, where the density operator $\rho$
is the {\sl quantum state} of the system \cite{2}.
\par
Given a quantum state $\rho$, by {\it support} of $\rho$ we mean any
{\it concrete} subset ${\mathcal S}(\rho)$ of specimens of the
physical system \cite{3}, whose quantum state is
$\rho$. Given a support ${\mathcal S}(\rho)$, by $\bf A({\mathcal
S}(\rho))$ we denote the concrete subset of all specimens in
${\mathcal S}(\rho)$ which {\sl actually} undergo a measurement of
$A$. In the following, we shall write simply $\bf A$ instead of $\bf
A({\mathcal S}(\rho))$ to avoid a cumbersome notation, whenever no
confusion is likely.
\par
By {\sl elementary} observable we mean any observable $E$
having only $0$ or $1$ as possible outcomes, and hence
represented by a projection operator $\hat E$; the
expected value $Tr(\rho\hat E)$ of an elementary observable $E$ coincides with the
{\it probability} that outcome $1$ occurs in a measurement of $E$.
By $\mathcal E$ we denote the set
of all elementary observables, and by $\hat{\mathcal E}({\mathcal H})$ the set of all projection
operators of $\mathcal H$.
\par
Fixed any support ${\mathcal S}(\rho)$, in correspondence with every
elementary observable $E$ we define the following extensions of $E$
in ${\mathcal S}(\rho)$.
\begin{description}
\item[-]
the set ${\bf E}$ of the specimens in ${\mathcal S}(\rho)$ which
{\it actually} undergo a measurement of $E$;
\item[-]
the set ${\bf E}_1\subseteq{\bf E}$ (resp., ${\bf E}_0\subseteq{\bf E}$) for which the
outcome $1$ (resp., 0) of $E$ has been obtained.\par
\end{description}
In agreement with Quantum Mechanics, we assume that the following statements hold
\cite{3}.
\begin{description}
\item[($2.1.i$)]
If $E$ is an elementary observable, then for every $\rho$ a support
${\mathcal S}(\rho)$ exists  such that ${\bf E}\neq\emptyset$.
\item[($2.1.ii$)]
For every support ${\mathcal S}(\rho)$, ${\bf E}_1\cap{\bf
E}_0=\emptyset$ and ${\bf E}_1\cup{\bf E}_0={\bf E}$, for every
$\rho$.
\item[($2.1.iii$)]
If $Tr(\rho \hat E)\neq 0$ then a support ${\mathcal S}(\rho)$
exists such that ${\bf E}_1\neq\emptyset$, and
\item[\qquad] \quad\;
if $Tr(\rho \hat E)\neq 1$, then a support ${\mathcal S}(\rho)$
exists such that ${\bf E}_0\neq\emptyset$.
\end{description}
In Quantum Theory \cite{2},
if $\hat B=f(\hat A)$ holds for two
self-adjoint operators
$\hat A$ and $\hat B$, then a measurement
of the observable $B$, henceforth denoted by $f(A)$, can be performed
by measuring $A$ and then transforming the outcome $a$ by the function $f$ into the outcome $b=f(a)$ of $B$.
As a consequence, the following statements hold in Quantum Theory.
$$
\hbox{If}\quad \hat B=f(\hat A)\quad\hbox{then}\quad x\in{\bf A}\hbox{ implies }x\in{\bf B}.\eqno{(2.2)}
$$
If $[\hat A,\hat B]={\bf 0}$,
then a third self-adjoint operator $\hat C$ and two functions $f$ and $g$ exist so that
$\hat A=f(\hat C)$ and $\hat B=g(\hat C)$ \cite{2}. Therefore,
$A$ and $B$ can be measured together if the
corresponding operators commute with each other.
Thus, the following implications hold.
\begin{description}
\item[($2.1.iv$)] $\{E_j\}\subseteq{\mathcal E}$ and
$ [\hat E_j, \hat E_k]={\bf 0}$ $\forall j,k$\quad imply \quad
$\forall\rho$ $\exists{\mathcal S}(\rho)$  such that $\cap_j{\bf
E_j}\neq\emptyset$.
\item[($2.1.v$)]
\; If $[\hat A,\hat B]={\bf 0}$ and $\hat D=f(\hat A,\hat B)$
then $x\in{\bf A}\cap{\bf B}$ implies $x\in{\bf D}$, $\forall {\mathcal S}(\rho)$.
\item[($2.1.vi$)]
If $F,G\in{\mathcal E}$ and $\hat F\hat G={\bf 0}$,
i.e. if $\hat F\bot \hat G$,
then ${\bf F}_1\cap{\bf G}_1=\emptyset$, $\forall
{\mathcal S}(\rho)$, $\forall\rho$,
\item[\qquad\quad] i.e.
in every simultaneous measurement of $F$ and $G$ the outcome $1$ for $F$ and $1$ for $G$
\item[\qquad\quad]  are mutually
exclusive; in this case the projection $\hat E =\hat F+\hat G$ belongs to $\hat{\mathcal E}({\mathcal H})$;
\item[\qquad\quad] conversely,
if $\hat F\bot G$ then $\hat E =\hat F+\hat G$ represents an elementary observable $E$ whose
\item[\qquad\quad]
measurement's outcome can be the sum of the simultaneous outcomes of $F$ and $G$.
\end{description}
%%%%%%%%%%%%%%%%%%%%%%%%%%%%%%%%%%%%%%%%%%%%%%%%%%%%%%%%%%%%%%%%%%%%%%%%%%%%%%%%%%
\subsection{Evaluations of elementary observables in Quantum Theory}
In general, given two elementary observables $E$ and $T$ we say
that $E$ can be {\sl evaluated} by $T$ if, according to Quantum Mechanics, the following perfect correlation
holds:
\vskip.4pc\noindent{\sl ``the
outcome of \,$T$ is $1$ if and only if the outcome of $E$ is $1$ in every
simultaneous measurement''.}\vskip.4pc\noindent
By
{\sl evaluation} of $E$ by $T$ we mean to assign $E$ the value obtained as
outcome of an actual measurement of $T$. Then we can give the following formal definition.
%%%%%%%%%%%%%%%%%%%%%%%%%%%%%%%%%%%%%%%%%%%%%%%%%%%%%%%%%%
\vskip1pc\noindent {\bf Definition
2.1.} {\sl Given $E,T\in\mathcal E$, the elementary observable $E$ can be evaluated
by $T$ when the system is assigned the state $\rho$, written $E\prec\rho\succ T$, if
\begin{description}
\item[{\rm (i)}] a support ${\mathcal S}(\rho)$ exists such that ${\bf T}\cap {\bf E}\neq\emptyset$ (simultaneous measurability),
\item[{\rm (ii)}] if $x\in{\bf E}\cap{\bf T}$ \quad then
[$x\in{\bf T}_1$ if and only if $x\in{\bf E}_1$] and [$x\in{\bf T}_0$ if and only if
$x\in{\bf E}_0$],
$\forall{\mathcal S}(\rho)$.
\end{description}}
\noindent
If an evaluation of $E$ by $T$ were identifiable with a measurement of $E$ in the state $\rho$, then
${\bf T}_1\subseteq{\bf
E}_1$ and ${\bf T}_0\subseteq{\bf E}_0$ should hold for any ${\mathcal S}(\rho)$. But the relation $E\prec\rho\succ T$ is symmetric;
thus the identification would be fully expressed by the following statement.
$$
E\prec\rho\succ T\qquad\hbox{if and only if}\qquad {\bf T}_1={\bf
E}_1\;\hbox{and}\; {\bf T}_0={\bf E}_0\,,\;\forall{\mathcal
S}(\rho). \eqno(2.3)
$$
Another relation $E\quad^\rho\lrcorr T$ can be defined as follows
\vskip.5pc\noindent
{\bf Definition 2.2.}
{\sl The relation $T\quad^\rho\lrcorr E$ holds if \quad $[\hat T,\hat E]={\bf 0}$ and $\hat T\rho=\hat E\rho$ hold.}
\vskip.5pc\noindent Relation $\quad^\rho\lrcorr$ is stronger than $\prec\rho\succ$; indeed the following proposition holds.
\vskip.5pc\noindent
{\bf Proposition 2.1.} {\sl Given $E,T\in\mathcal E$,
if $E\quad^\rho\lrcorr T$ holds then $E\prec\rho\succ T$ holds too.}
\vskip.5pc\noindent
{\bf Proof.}
Condition (i) in def.2.1 follows from (2.1.iv). Then condition (ii) holds
if the probability of the pairs $(1,0)$ and $(0,1)$ in a simultaneous measurement of $T$ and $F$ is zero, i.e.,
if $Tr(\rho\hat T[{\bf
1}-\hat E])={\bf 0}$ and $Tr(\rho[{\bf 1}-\hat T]\hat E)={\bf 0}$, i.e. if
$\hat T\rho=\hat T\hat E\rho$ and $\hat
E\rho=\hat E\hat T\rho$.
\subsection{Evaluations are not measurements}
Now
we shall single out seven elementary observables
$E^\alpha,E^\beta,F,G^\alpha,G^\beta,L^\alpha,L^\beta$
of a particular quantum system,
chosen so that they are all measurable together {\sl if the identification (2.3)
of evaluations with measurements holds}.
Then we show that their simultaneous outcomes
$\eta^\alpha,\eta^\beta,\phi,\gamma^\alpha,\gamma^\beta,\lambda^\alpha,\lambda^\beta$
must satisfy the following constraints, where $f$ is the function $f(\xi)=2\xi-1$.
$$
\left\{
\begin{array}{llll}
\textrm{   i)}\quad &f(\eta^\alpha)f(\phi)&=-f(\gamma^\alpha)f(\lambda^\alpha),\\
\textrm{  ii)}\quad &f(\eta^\beta)f(\phi)&=-f(\gamma^\beta)f(\lambda^\alpha),\\
\textrm{ iii)}\quad &f(\eta^\beta)f(\phi)&=-f(\gamma^\alpha)f(\lambda^\beta),\\
\textrm{  iv)}\quad &f(\eta^\alpha)f(\phi)&=f(\gamma^\beta)f(\lambda^\beta),\\
\end{array}\right.
\eqno(2.4)
$$
Since each factor $f(\xi)$ in (2.4) must be $-1$ or $+1$,
these constraints are contradictory, because, by elementary algebra, they
imply $f(\gamma^\alpha)f(\gamma^\beta)=-f(\gamma^\alpha)f(\gamma^\beta)$.
Thus, in Quantum Mechanics evaluations cannot be identified with measurements.
\vskip.5pc
To realize such a program,
we consider a quantum system described in the Hilbert space
${\mathcal H}={\mathcal H}_1\otimes{\mathcal H}_2\otimes{\mathcal H}_3\otimes{\mathcal H}_4$,
where each ${\mathcal H}_k$ is ${\bf C}^2$. The following projection operators represent
seven elementary
observables\footnote{Our argument makes use of the mathematical
setting adopted by Greenberger, Horne, Shimony and Zeilinger (GHSZ)
\cite{4} to prove that a given
set of premises are in contradiction with Quantum Theory.
In \cite{3} we proved that if the premises of GHSZ theorem
are modified, then the contradiction does not necessarily arise.
The hypotheses of the present argument, i.e.
identification (2.3), are different from the premises of GHSZ.
Thus the occurrence of a contradiction needs an explicit proof.}.
\par
$\hat E^\alpha=\frac{1}{2}\left[\begin{array}{cc}
1 & 1 \\
1 & 1 \\
\end{array}\right]_1\otimes{\bf 1}_2\otimes{\bf 1}_3\otimes{\bf 1}_4$;\qquad
$\hat E^\beta=\frac{1}{2}\left[\begin{array}{cc}
1 & -i \\
i & 1 \\
\end{array}\right]_1\otimes{\bf 1}_2\otimes{\bf 1}_3\otimes{\bf 1}_4$;
\par
$\hat F={\bf 1}_1\otimes\frac{1}{2}\left[\begin{array}{cc}
1 & 1 \\
1 & 1 \\
\end{array}\right]_2\otimes{\bf 1}_3\otimes{\bf 1}_4$;\par
$\hat G^\alpha={\bf 1}_1\otimes{\bf 1}_2\otimes\frac{1}{2}\left[\begin{array}{cc}
1 & 1 \\
1 & 1 \\
\end{array}\right]_3\otimes{\bf 1}_4$;\qquad
$\hat G^\beta={\bf 1}_1\otimes{\bf 1}_2\otimes\frac{1}{2}\left[\begin{array}{cc}
1 & -i \\
i & 1 \\
\end{array}\right]_3\otimes{\bf 1}_4$;\par
$\hat L^\alpha={\bf 1}_1\otimes{\bf 1}_2\otimes{\bf 1}_3\otimes\frac{1}{2}\left[\begin{array}{cc}
1 & 1 \\
1 & 1 \\
\end{array}\right]_4$;\qquad
$\hat L^\beta={\bf 1}_1\otimes{\bf 1}_2\otimes{\bf 1}_3\otimes\frac{1}{2}\left[\begin{array}{cc}
1 & -i \\
i & 1 \\
\end{array}\right]_4$;
\vskip.5pc\noindent
Let the physical system be assigned the pure state $\rho_0=\vert\psi_0\rangle\langle\psi_0\vert$,
where
$$
\psi_0=\frac{1}{\sqrt{2}}\left(\left[\begin{array}{c}
1  \\0
\end{array}\right]_1
\otimes\left[\begin{array}{c}
1  \\0
\end{array}\right]_2
\otimes\left[\begin{array}{c}
0  \\1
\end{array}\right]_3
\otimes\left[\begin{array}{c}
0  \\1
\end{array}\right]_4
-\left[\begin{array}{c}
0  \\1
\end{array}\right]_1
\otimes\left[\begin{array}{c}
0  \\1
\end{array}\right]_2
\otimes\left[\begin{array}{c}
1  \\0
\end{array}\right]_3
\otimes\left[\begin{array}{c}
1  \\0
\end{array}\right]_4
\right).
$$
The four projection operators $\hat E^\alpha,\hat F,\hat G^\beta,\hat L^\alpha$ commute with
each other; hence, by (2.1.iv), all the corresponding elementary observables can be measured together, i.e.
a support ${\mathcal S}(\rho_0)$ and a specimen $x_0\in{\mathcal S}(\rho_0)$ exist such that
$$
x_0\in {\bf E}^\alpha\cap{\bf F}\cap{\bf G}^\beta\cap{\bf L}^\alpha.
$$
Let $\eta^\alpha,\phi,\gamma^\beta,\lambda^\alpha$ be their respective
outcomes relative to $x_0$. Now, the projection operator
$$
\hat M=\frac{{\bf 1}-f(\hat E^\alpha)f(\hat F)f(\hat L^\alpha)}{2}
$$
is a function of $\hat E^\alpha,\hat F,\hat L^\alpha$; therefore $x_0\in{\bf M}$ because of (2.1.v)
and $\mu=\frac{1}{2}[1-f(\eta^\alpha)f(\phi)f(\lambda^\alpha)]$ must be the outcome of the elementary observable $M$
so measured on $x_0$.
\par
But
$[\hat M,\hat G^\alpha]={\bf 0}$ trivially holds; moreover, a direct calculation shows that
the equation $\hat M\rho_0=\hat G^\alpha\rho_0$ is satisfied; then
$M\;^{\rho_0}\lrcorr G^\alpha$ holds and Prop.2.1  implies $M\prec\rho_0\succ G^\alpha$. If
the identification (2.3) holds, then
$x_0$ belongs to ${\bf G}^\alpha$ and $\gamma^\alpha=\mu=\frac{1}{2}[1-f(\eta^\alpha)f(\phi)f(\lambda^\alpha)]$ is to be identified
as the outcome of the measurement of $G^\alpha$ on $x_0$. Then, from
$f(\gamma^\alpha)\equiv 2\gamma^\alpha-1$ we find (2.4.i). Then,
$x_0\in{\bf E}^\alpha\cap{\bf F}\cap{\bf G}^\alpha\cap{\bf G^\beta}\cap{\bf L^\alpha}$ and the
constraint
(2.4.i) must hold.
\par
Now we derive (2.4.ii).
By defining $\hat N=\frac{{\bf 1}-f(\hat F)f(\hat G^\beta)f(\hat L^\alpha)}{2}$,
we can verify that  $N\;^{\rho_0}\lrcorr E^\beta$.
Then,
following the argument which led us to (2.4.i), we obtain that
$x_0\in{\bf E}^\alpha\cap{\bf E}^\beta\cap{\bf F}\cap{\bf G}^\alpha\cap{\bf G^\beta}\cap{\bf L^\alpha}$
and (2.4.ii) holds.
\par
Similarly,
by defining $\hat R=\frac{{\bf 1}-f(\hat E^\beta)f(\hat F)f(\hat G^\alpha)}{2}$,
it turns out that  $R\;^{\rho_0}\lrcorr L^\beta$ holds.
Then,
we can imply that
$x_0\in{\bf E}^\alpha\cap{\bf E}^\beta\cap{\bf F}\cap{\bf G}^\alpha\cap{\bf G^\beta}\cap{\bf L^\alpha}\cap{\bf L}^\beta$
and (2.4.iii) hold.
\par
But we can also define
$\hat S=\frac{{\bf 1}+f(\hat E^\alpha)f(\hat F)f(\hat G^\beta)}{2}$. The elementary observable $S$
turns out to satisfy the relation $S\;^{\rho_0}\lrcorr L^\beta$, which implies (2.4.iv).
\par
Then all the
constraints
(2.4) must hold for the simultaneous measurement of
$E^\alpha,E^\beta,F,G^\alpha,G^\beta,L^\alpha,L^\beta$
on the specimen $x_0$.
Thus,
identification (2.3) cannot hold in Quantum Mechanics.
\section{Evaluations as simulations of measurements}
The impossibility of identifying evaluations with measurements, proved in section 2,
opens the question of establishing the conceptual status of the evaluations in Quantum Theory;
since evaluations are diffusely practiced in Quantum Physics, it is important give an answer
to such a question. In this section we show that an evaluation of $E$ by $T$ works as a
{\it perfect simulation} of a measurement of $E$.
\vskip.5pc
A way to understand
which is the conceptual status of an
evaluation of $E$ by $T$ is to compare the
{\sl physical implications} of the occurrences of the outcomes of
$E$ with the physical implications of the occurrences of the
corresponding outcomes of $T$.
Such a comparison amounts to compare the conditional probabilities
$P(F\mid E)=Tr(\rho \hat F\hat
E)/Tr(\rho \hat E)$, $P(F\mid E')=Tr(\rho \hat F\hat
E')/Tr(\rho \hat E')$ established by Quantum Theory with the
conditional probabilities
$P(F\mid T)=Tr(\rho \hat F\hat
T)/Tr(\rho \hat T)$, $P(F\mid T')=Tr(\rho \hat F\hat
T')/Tr(\rho \hat T')$, where $E'\equiv 1-E$ and $T'\equiv 1-T$.
\par
These conditional probabilities are defined whenever $[\hat F,\hat
T]=[\hat F,\hat E]={\bf 0}$; therefore the domain of the comparison
is
${\mathcal F}(E,T)=\{F\in {\mathcal E}\mid  [\hat F,\hat T]=[\hat
F,\hat E]={\bf 0}\}$.
The following statement follows
from prop. 2.1.
$$
\hbox{If}\;\; T\quad^{\rho_0}\lrcorr E\quad\hbox{then}\;\;
P(F\mid T)=\frac{Tr(\rho\hat F\hat T)}{Tr(\rho\hat
T)}=\frac{Tr(\rho\hat F\hat E)}{Tr(\rho\hat E)}= P(F\mid E), \;
\forall F\in{\mathcal F}(E,T).\eqno(3.1)
$$
Now, from Def.2.2 it follows that $T\quad^\rho\lrcorr E$,
holds iff $T'\quad^\rho\lrcorr E'$; so
(3.1) extends to
$$
\hbox{If}\;\; T\quad^{\rho_0}\lrcorr E\quad\hbox{then}\;\;
P(F\mid T')=\frac{Tr(\rho\hat F{\hat T}')}{Tr(\rho{\hat
T}')}=\frac{Tr(\rho\hat F{\hat E}')}{Tr(\rho{\hat E}')}= P(F\mid
E'), \; \forall F\in{\mathcal F}(E,T).\eqno(3.2)
$$
Therefore, in the case that $T\quad^{\rho_0}\lrcorr E$ holds,
the effects of an evaluation of $E$ by $T$
are indistinguishable from the effects of the
occurrence of the same outcome in a direct measurement of $E$.
In other words, the measurement of $E$ is perfectly {\sl
simulated} by a measurement of a evaluating observable $T$.
\vskip.5pc\noindent{\bf Remark 3.1.}
In fact, once fixed the evaluated observables $E$,
our results are obtained for evaluating observables $T$ such that
$T\quad^\rho\lrcorr E$; these observables form a subset
of all evaluating observables evaluating $E$ which have to satisfy the weaker condition
$T\prec\rho\succ E$. However, the converse of Prop.2.1,
i.e. the implication $T\prec\rho\succ E$ implies $T\quad^\rho\lrcorr E$,
immediately follows from the principle of Quantum Mechanics which establishes that
$T$ and $E$ are measurable together if and only if $[\hat T,\hat E]={\bf 0}$.
Therefore, if such a principle is assumed, our results holds for all evaluating observables.
%%%%%%%%%%%%%%%%%%%%%%%%%%%%%%%%%%%%%%%%%%%%%%%%%%%%%%%%%%%%%%%%%%%%%%%%%%%%%%%%%%%%%%%%%%%%%%%%%%%%%%
\section{To Evaluate $E$ while incompatible observables are measured}
Let $T$, $E$ and $F$ be elementary observables such that
$T\quad^\rho\lrcorr E$ and $[\hat T,\hat F]={\bf 0}$, so that $T$
can be measured together with $F$; in the case in which
$F\notin{\mathcal F}(E,T)$, i.e. if $[\hat
F,\hat E]\neq{\bf 0}$, a measurement of $T$ simultaneous to a
measurement of $F$ cannot be interpreted as a simulation of a
measurement of $E$, according to section 3.1, because there is no conditional
probability $P(F\mid E)$ to be compared with $P(F\mid T)$.
\par
The double slit experiment described in the introduction is an emblematic
example of this circumstance. In that case it is possible to evaluate the
which slit observable $W$ by $T_W$; but if also the elementary observable $F(\Delta)$,
which indicates the localization in the region $\Delta$ of the final screen, is measured,
then the evaluation cannot be interpreted as simulation of a measurement of $W$ because
$P(F(\Delta)\mid W)$ does not exist since
$[\hat W,\hat F(\Delta)]\neq{\bf 0}$.
\par
In the present section we address the interpretative lack which occurs in a
general situation where
$$T\quad^\rho\lrcorr E,
\quad F\in{\mathcal F}(T)\equiv\{F\in{\mathcal E}\mid\quad[\hat T,\hat F]={\bf 0}\},\quad\hbox{but}\quad [\hat E,\hat F]\neq{\bf 0},
\eqno(4.1)
$$
and $E$ is evaluated by $T$ simultaneously to a measurement of $F$.
\par
In section 4.1 we establish results which allow us to provide the problem with an answer
we formulate in section 4.2. Remark 3.1 applies also in this more general case.
\subsection{Consistency of assignment by evaluations}
Let us start with the following result of Cassinelli and Zangh\`\i\/ \cite{5}.
\vskip.5pc\noindent
{\bf Lemma 4.1.}
{\sl Let $\hat{\mathcal A}$ be a Von Neumann
algebra\footnote{A Von Neumann algebra \cite{6}
is a subset $\hat{\mathcal A}$ of bounded linear operators of the Hilbert space $\mathcal H$
such that $\hat{\mathcal A}=(\hat{\mathcal A}^\prime)^\prime\equiv\hat{\mathcal A}^{\prime\prime}$, where $\hat{\mathcal A}'$ denotes the
{\sl commutant} of $\hat{\mathcal A}$, i.e. the set of all bounded linear operators $\hat B$ of $\mathcal H$
such that $[\hat B,\hat A]={\bf 0}$ for all $\hat A\in\hat{\mathcal A}$.
The theory of Von Neumann algebras \cite{6} shows that if $\hat\Pi(\hat{\mathcal A})$ is the set
of all projection operators in the Von Neumann algebra $\hat{\mathcal A}$,
then $\hat{\mathcal A}=\hat\Pi(\hat{\mathcal A})^{\prime\prime}$.
}, and let $\Pi(\hat{\mathcal A})$ be the set of all projection operators in $\hat{\mathcal A}$.
Given a projection operator $\hat E\in\Pi(\hat{\mathcal A})$, for every density operator $\rho$
the function
$$
p_\rho(E\& \;:  \Pi(\hat{\mathcal A})\to [0,1],\quad p_\rho(E\& F)=Tr(\rho\hat E\hat F\hat E)
$$
is the unique functional which satisfies the following conditions.
\begin{description}
\item[{\rm (i)}]
If $\hat F\in\Pi(\hat{\mathcal A})$ and $[\hat F,\hat E]={\bf 0}$ then $p_\rho(E\& F)=Tr(\rho\hat E\hat F)$;
\item[{\rm (ii)}]
if $\{\hat F_j\}_{j\in J}\subseteq \Pi(\hat{\mathcal A})$ is any countable
family of projection operators such that $\sum_{j\in J}\hat F_j\equiv \hat F\in\Pi(\hat{\mathcal A})$,
then \quad$p_\rho( E\& F)=\sum_{j\in J}p_\rho( E\&
F_j)$.
\end{description}
}\vskip.4pc\noindent
Since the set $\hat{\mathcal F}(\hat T)=\{\hat F\in{\mathcal E}({\mathcal H})\mid [\hat F,\hat T]={\bf 0}\}$
is just the set of all projection operators of the Von Neumann algebra $\hat{\mathcal A}(\hat{T})=\{\hat T\}'$,
and hence $\hat{\mathcal F}(\hat T)$ generates $\hat{\mathcal A}(\hat{T})$, the following theorem
can be proved by means of Lemma 4.1.
\vskip.5pc\noindent
{\bf Theorem 4.1.}
{\sl Let $T$ be an elementary observable. Given $E\in{\mathcal F}(T)$,  for every quantum state
$\rho$ the mappings
\par
$p_\rho(E\&\;:{\mathcal F}(T)\to[0,1]$, $p_\rho(E\& F)=Tr(\rho\hat E\hat F\hat E)$, and\par
$p_\rho(E'\&\;:{\mathcal F}(T)\to[0,1]$, $p_\rho(E'\& F)=Tr(\rho\hat E'\hat F\hat E')$
\par\noindent
are the unique functionals which satisfy the following conditions.
\begin{description}
\item[{\rm (C.1)}]
If $F\in{\mathcal F}(T)$ and $[\hat F,\hat E]={\bf 0}$ then $p_\rho(E\& F)=Tr(\rho\hat E\hat F)$ and $p_\rho(E'\& F)=Tr(\rho\hat E'\hat F)$;
\item[{\rm (C.2)}]
if $\{F_j\}_{j\in J}\subseteq {\mathcal F}(T)$ is any countable
family such that $\sum_{j\in J}\hat F_j\equiv \hat F\in\hat{\mathcal
F}(\hat T)$, then \quad$p_\rho( E\& F)=\sum_{j\in J}p_\rho( E\&
F_j)$ and $p_\rho( E'\& F)=\sum_{j\in J}p_\rho( E'\& F_j)$.
\end{description}}\vskip.4pc\noindent
Theorem 4.1 implies that if $E$ can be assigned values which extend its measured values to the case
that $E$ is not measured,
in a way which is logically consistent with the
outcomes of whatever actually measured observable $F\in{\mathcal F}(T)$, then it is necessary that such an assignment be
ruled over by the probability
$$
p_\rho(E\& F)=Tr(\rho\hat E\hat F\hat E)\quad {\rm (resp.,}\;p_\rho(E'\& F)=Tr(\rho\hat E'\hat F\hat E'){\rm )}
$$
of assigning $E$ value 1 (resp., 0) and obtaining value 1 for $F$.
\par
However, the agreement with probability $P_\rho$ is not sufficient. Logical consistency requires
that the condition must hold too.
$$
p_\rho(E'\& F)+p_\rho(E\& F)=Tr(\rho\hat F)\quad\hbox{for all }F\in{\mathcal F}(T)\leqno{\rm (C.3)}
$$
In general, (C.3) does not hold. Indeed, we have\par
$Tr(\rho\hat F)=Tr(\rho[\hat E+\hat E']\hat F[\hat E+\hat E'])
=p_\rho(E\& F)+p_\rho(E'\& F)+2{\bf Re}Tr(\hat E\hat F\hat E')$.
\par\noindent
Therefore, if $[\hat F,\hat E]\neq{\bf 0}$ the presence of a non-vanishing {\sl interference} term
$2{\bf Re}Tr((\hat E\hat F\hat E')$ cannot be excluded, and (C.3) is violated.
\par
The following theorem proves that to assign $E$ just the outcome of the evaluation of $E$ by $T$
guarantees all conditions (C.1), (C.2), (C.3).
\vskip.5pc\noindent
{\bf Theorem 4.2.}
{\sl Given two elementary observables $E,T\in{\mathcal E}$, if $E\quad^\rho\lrcorr T$,
so that $T$ can evaluate $E$,
then for all $F\in{\mathcal F}(T)$ the equalities $p_\rho(E\& F)=Tr(\rho\hat T\hat F)$
and $p_\rho(E'\& F)=Tr(\rho\hat T'\hat F)$ hold.
Furthermore, also (C.3) holds.}
\vskip.5pc\noindent
{\bf Proof.}
If $E\quad^\rho\lrcorr T$, then $\hat E\rho=\hat T\rho$ and $\rho\hat E=\rho \hat T$ hold;
therefore $p_\rho (E\& F)=Tr(\rho\hat E\hat F\hat E)=Tr(\hat E\rho\hat E\hat F)=Tr(\hat T\rho\hat E\hat F)
=Tr(\hat T\rho\hat T\hat F)=Tr(\rho\hat T\hat F)$ holds because $[\hat F,\hat T]={\bf 0}$. Analogously, since
$E\quad^\rho\lrcorr T$ implies $E'\quad^\rho\lrcorr T'$, $p_\rho(E'\& F)=Tr(\rho\hat T'\hat F)$ is proved.
Then $p_\rho(E\& F)+p_\rho(E'\& F)=Tr(\rho\hat T\hat F)+Tr(\rho\hat T'\hat F)=Tr(\rho\hat F)$; thus also (C.3) is proved.
\subsection{Conclusions}
In virtue of theorem 4.2 we can state the following conclusions, which are valid whenever $E\quad^\rho\lrcorr T$.
\begin{description}
\item[{\rm (Ev.1)}]
It is logically consistent to assign $E$ the value obtained by an evaluation by $T$ together with simultaneous measurement of
whatsever family $\{F_j\}$ of observables from ${\mathcal F}(T)$, also if $[\hat E,\hat F]\neq{\bf 0}$.
\item[{\rm (Ev.2)}]
The probability which rules over such an assignment and the occurrences of the outcomes
extends the probability established by Quantum Theory for actually measured outcomes.
\item[{\rm (Ev.3)}]
The consistency is guaranteed within the domain ${\mathcal F}(T)\subseteq {\mathcal E}$, not elsewhere.
\end{description}
According to these conclusions, to assign $E$ the value evaluated by $T$ cannot provoke contradiction if the evaluation
is performed together with whatsever measurement of an observable $F$, also if $[\hat F,\hat E]\neq{\bf 0}$. Thus,
our conclusions (Ev) contribute to fulfill the interpretative lack about evaluations which
arises in the circumstances where conditions (4.1) hold.
\vskip.5pc
As further implications of our results, we can deduce that it is possible
to exploit evaluations for consistently assigning non-commuting
observables simultaneous values. Indeed, let us consider two elementary observables $E_1$, $E_2$ such that
$E_1\quad^\rho\lrcorr T_1$ and $E_2\quad^\rho\lrcorr T_2$,
so that they can be evaluated by $T_1$ and $T_2$, respectively, in the state $\rho$.
If $[\hat T_1,\hat T_2]={\bf 0}$ and $E_1,E_2\in{\mathcal F}(T_1)\cap{\mathcal F}(T_2)$,
the evaluations of both $E_1$ and $E_2$ can be
obtained simultaneously by measuring together $T_1$ and $T_2$. Therefore $E_1$ and $E_2$ can be consistently assigned
simultaneous values, also if $[\hat E_1,\hat E_2]\neq{\bf 0}$.
Meaningful examples are developed in \cite{1},\cite{7},\cite{8}.
\vskip.5pc
These results do no conflict with the impossibility of identifying evaluations with measurement
proved in section 2.3; on the contrary, they explain it.
Therein the observables actually measured are $E^\alpha,F,G^\beta,L^\alpha$. Hence, also
the elementary observables $M$ and $N$ represented by the projection operators
$\hat M=\frac{{\bf 1}-f(\hat E^\alpha)f(\hat F)f(\hat L^\alpha)}{2}$ and
$\hat N=\frac{{\bf 1}-f(\hat F)f(\hat G^\beta)f(\hat L^\alpha)}{2}$ are actually measured according to (2.1.v).
Since $\hat N=\frac{{\bf 1}-f(\hat F)f(\hat G^\beta)f(\hat L^\alpha)}{2}$,
$N\quad^{\rho_0}\lrcorr E^\beta$ and $G^\alpha,E^\beta\in{\mathcal F}(M)\cap{\mathcal F}(N)$ hold,
$G^\alpha$ and $E^\beta$ can be consistently assigned values.
However, $L^\beta$ cannot be consistently assigned the value
evaluated by $R$, though $R\quad^{\rho_0}\lrcorr L^\beta$, because $R$ is not actually measured.
Indeed, since $\hat R=\frac{{\bf 1}-f(\hat E^\beta)f(\hat F)f(\hat G^\alpha)}{2}$, the measurement of $R$
entails the measurement of $E^\beta$ and $G^\alpha$; but
$[\hat E^\beta,\hat E^\alpha]\neq{\bf 0}$, $[\hat G^\alpha,\hat G^\beta]\neq{\bf 0}$
and $E^\alpha$, $G^\beta$ are actually measured.
Thus, according to our results, a consistent value assignment also to $E^\beta$ is not guaranteed,
and indeed an inconsistency arises in attempting it.
%%%%%%%%%%%%%%%%%%%%%%%%%%%%%%%%%%%%%%%%%%%%%%%%%%%%%%%%%%%%%%%%%%%%%%%%%%%


\begin{thebibliography}{20}
\bibitem{1}
Nistic\`o G 2008: {\it J. Phys. A: Math. Theor.} {\bf 41} 125302
\bibitem{2}
Von Neumann J 1955 {\it Mathematical Foundations of Quantum Mechanics}
(Princeton: Princeton University Press)
\bibitem{3}
Nistic\`o G and Sestito A 2011: {\it Found.Phys.} {\bf 41} 1263
\bibitem{4}
Greenberger D M,
Horne M A, Shimony A and Zeilinger A 1990: {\it Am.J.Phys.} {\bf 58} 1131
\bibitem{5}
Cassinelli G and Zangh\`i\/ N 1983: {\it Il Nuovo Cimento B} {\bf 73} 237
\bibitem{6}
Dixmier J 1957: {\it Les algebres d'op\'erateurs dans l'espace
Hilbertien} (Paris: Gauthier-Villars)
\bibitem{7}
Sestito A 2008: {\it Found.Phys.} {\bf 38} 935
\bibitem{8}
Nistic\`o G and Sestito A 2012: {\it International Journal of Quantum
Information} {\bf 10} 1250055
\end{thebibliography}
\end{document}